\documentclass[a4paper, 12pt, openany, oneside]{article}
\usepackage[cp1251]{inputenc} 
\usepackage[english]{babel} 
\usepackage{ graphics,amsfonts, amsmath,bm,amssymb, array}
\usepackage[dvips]{graphicx}
\renewcommand{\@biblabel}[1]{#1.\hfill}

\newcommand{\diag}{\rm \diag\, }

\renewcommand{\Re}{\mathop{\rm Re\,}}
\renewcommand{\Im}{\mathop{\rm Im\,}}

\makeatother {

\begin{document}
\thispagestyle{empty} \large
\renewcommand{\abstractname}{Abstract}
\thispagestyle{empty}
\renewcommand{\refname}{\begin{center} REFERENCES\end{center}}

 \begin{center}
\bf Surface plasmons in metallic films of arbitrary thickness with mirror boundary conditions
\end{center}\medskip
\begin{center}
  \bf A. V. Latyshev\footnote{$avlatyshev@mail.ru$} and
  A. A. Yushkanov\footnote{$yushkanov@inbox.ru$}
\end{center}\medskip

\begin{center}
{\it Faculty of Physics and Mathematics,\\ Moscow State Regional
University, 105005,\\ Moscow, Radio str., 10--A}
\end{center}\medskip

\begin{abstract}
A metallic film of arbitrary thickness is considered. We show that the problem
of description of surface plasma oscillations (surface plasmons) with reflection
boundary conditions allows analytic solution. Besides, this problem
allows generalization for more general case of conducting matter
(in particular, and for semiconductors).

\medskip

structures
\end{abstract}

\begin{center}
{\bf Introduction}
\end{center}

Problem about surface plasma oscillations long time draws to itself at\-ten\-tion
\cite{F69}--\cite{F09}. It is connected as with theoretical interest to this
problem, and with numerous practical appendices.
Thus the majority of researches is devoted research
surface plasma waves on border of two substances. At the same time
The big interest causes process of distribution of the surface
plasma waves in thin films, in particular, the metal.

Researches of interaction of an electromagnetic wave with
metal film were spent basically for a case
mirror dispersion of electrons on a film surface.
The macroscopical approach was thus used. The kinetic equations
thus were not used. In the present work it is shown, that for films
any thickness the problem with application of the kinetic approach supposes the analytical solution for mirror boundary conditions.

For more general boundary conditions a problem for a film of any thickness essentially becomes complicated and does not suppose generally the analytical solution.

Let's notice, that the most part of our reasonings will be fair for
more the general case conducting medium (in particular and semiconductor)
films.

\begin{center}
{\bf 1. The basic equations}
\end{center}

Let's consider a metal film of any thickness.
We take the Cartesian system of coordinates with the origin of coordinates
with an axis $x$, directed per\-pen\-di\-cu\-larly to the surface of a film.
Axis $z$ we will direct along a direction of distribution of the surface
electromagnetic wave. We will notice, that in this case a magnetic field
it is directed along an axis $y$.
The origin of coordinates we will place in the middle of a film.
Let's designate a thickness film through $d$.

Out of a film the electromagnetic field is described by the wave equations
$$
\dfrac{1}{c^2}\dfrac{\partial^2 {\bf E}}{\partial t^2}-\triangle {\bf
E}=0, \qquad
$$

$$
\dfrac{1}{c^2}\dfrac{\partial^2 {\bf H}}{\partial t^2}-\triangle {\bf
H}=0.
$$ \medskip

Here ${\bf E}=\{E_x,0,E_z\}$, ${\bf H}=\{0,H_y,0\}$ are vectors of
electric and magnetic fields strength, $c$ is the speed of light.

The solution of these equations decreasing on infinity, looks like
$$
{\bf E}=\left\{\begin{array}{l}{\bf E}_1e^{-i\omega t+\alpha x+ikz},\qquad
x<-d/2, \\
{\bf E}_2e^{-i\omega t-\alpha x+ikz},\qquad x>d/2,
\end{array}\right.
\eqno{(1a)}
$$
$$
{\bf H}=\left\{\begin{array}{l}{\bf H}_1e^{-i\omega t+\alpha x+ikz},\qquad
x<-d/2, \\
{\bf H}_2e^{-i\omega t-\alpha x+ikz},\qquad x>d/2.
\end{array}\right.
\eqno{(1b)}
$$

In the equations (1a) and (1b) $\omega $ is the frequency of a wave,
$k$ is the wave number.
The attenuation parameter $\alpha$ is connected with these quantity by equation
$$
\alpha=\sqrt{k^2-\dfrac{\omega^2}{c^2}}.
\eqno{(2)}
$$

Then behaviour of electric and magnetic
fields of a wave in a film it is described by the following system
of differential equations \cite {K}:

$$
\left\{\begin{array}{l}
\dfrac{dE_z}{dx}-ik E_x+\dfrac{i\omega}{c}H_y=0, \\ \\
\dfrac{i\omega}{c}E_x-ik H_y=\dfrac{4\pi}{c}j_x,\\ \\
\dfrac{dH_y}{dx}+\dfrac{i\omega}{c}E_z=\dfrac{4\pi}{c}j_z.
\end{array}\right.
\eqno{(3)}
$$

Here $\,\bf j$ is the current density.

Let's consider two cases of a configuration of external fields. In
the first case we will consider a symmetric configuration of
$y$--components of a magnetic field and $x$--components electric
fields, and an antisymmetric configuration of $z$--components
electric field, i.e.
$$
H_y\Big(-\dfrac{d}{2}\Big)=H_y\Big(\dfrac{d}{2}\Big),\quad
E_x\Big(-\dfrac{d}{2}\Big)=E_x\Big(\dfrac{d}{2}\Big),\quad
E_z\Big(-\dfrac{d}{2}\Big)=-E_z\Big(\dfrac{d}{2}\Big).
$$

In the second case we will consider an antisymmetric configuration of
$y$--components of a magnetic field and $x$--components electric
fields, and a symmetric configuration of $z$--components
electric field, i.e.
$$
H_y\Big(-\dfrac{d}{2}\Big)=-H_y\Big(\dfrac{d}{2}\Big),\quad
E_x\Big(-\dfrac{d}{2}\Big)=-E_x\Big(\dfrac{d}{2}\Big),\quad
E_z\Big(-\dfrac{d}{2}\Big)=E_z\Big(\dfrac{d}{2}\Big).
$$

\begin{center}
{\bf 2. Dielectric permeability and impedance}
\end{center}

The impedance in both cases is thus certain equally as follows
$$
Z_j=\dfrac{E_z(-0)}{H_y(-0)},\qquad j=1,2.
$$

Let's consider the case of mirror reflection of electrons from a film surface.
 Then for quantities $Z_j \, (j=1,2)$ we have the following equalities \cite{F69}
$$
Z_{j}=-\dfrac{2i}{W(d)\Omega}
\sum\limits_{n=-\infty}^{n=\infty}\dfrac{1}{Q^2}
\Bigg[\dfrac{Q_x^2}{\varepsilon_{tr}-Q^2/\Omega^2}+
\dfrac{Q_z^2}{\varepsilon_{l}}
\Bigg], \quad j=1,2.
\eqno{(4)}
$$

And for $Z^{(1)}$ summation is conducted on odd $n$, and for
$Z^{(2)}$ on even, $\varepsilon_{tr}$ and $\varepsilon_l$
are accordingly the transvercal and longitudinal dielectric
permeability of plasma, $ \Omega =\omega/\omega_p $,
$ \varepsilon =\nu/\omega_p $, $ \omega_p $ is the plasma (Langmuir)
frequency of plasma ocsillations, $\omega_p^2=4\pi e^2N/m$, $e$ and $m$ are
charge and mass of electron, $N$ is the concentration (numerical density)
of electrons, $W(d)=\omega_p d/c $, the wave vector looks like
$$
\mathbf{Q}=\dfrac{c}{\omega_p}\mathbf{q},\qquad
\mathbf{q}=\Big\{\dfrac{\pi n}{d},0,k\Big\},\quad
$$
$$
q=|\mathbf{q}|=\sqrt{\dfrac{\pi^2 n^2}{d^2}+k^2},
\qquad n=0,\pm 1,\pm 2,\cdots,
$$
$k$ is the dimensional wave number.

Let's introduce an unit vector
$$
\mathbf{e}=\dfrac{\mathbf{q}}{q}=
\Big\{\dfrac{Q_x}{Q},0,\dfrac{Q_z}{Q}\Big\}=
\dfrac{1}{q}\Big\{\dfrac{\pi n}{d},0,k\Big\}.
$$

Then the formula (4) will be copied in the form
$$
Z_{j}=-\dfrac{2i}{W(d)\Omega}
\sum\limits_{n=-\infty}^{n=\infty}
\Bigg[\dfrac{e_x^2}{\varepsilon_{tr}-Q^2/\Omega^2}+
\dfrac{e_z^2}{\varepsilon_{l}}
\Bigg], \quad j=1,2,
$$
where $
e_x=\dfrac{\pi n}{qd},\, e_z=\dfrac{k}{q},
$
or, in explicit form
$$
Z_{j}=-\dfrac{2i}{W(d)\Omega}
\sum\limits_{n=-\infty}^{n=\infty}\dfrac{1}{q^2}
\Bigg[\dfrac{(\pi n/d)^2}{\varepsilon_{tr}-(cq/\Omega\omega_p)^2}
+\dfrac{k^2}{\varepsilon_{l}}\Bigg], \quad j=1,2,
\eqno{(5)}
$$
where $$
W(d)=\dfrac{\omega_p}{c}d.
$$

Further we introduce the quantity
$$
q_1=\dfrac{v_F}{\omega_p}q=\dfrac{v_F}{\omega_p}
\sqrt{\dfrac{\pi^2 n^2}{d^2}+k^2}.
$$

By means of this quantity the transvercal conductivity is calculated by
the formula
$$
\varepsilon_{tr}=1-\dfrac{3}{4\Omega q_1^3}\Bigg[
2(\Omega+i\varepsilon)q_1+
\Big[(\Omega+i\varepsilon)^2-q_1^2\Big]
\ln\dfrac{\Omega+i\varepsilon-q_1}
{\Omega+i\varepsilon+q_1}\Bigg],
$$
or $$
\varepsilon_{tr}=1-\dfrac{3}{4\Omega q_1}
\Bigg\{2\dfrac{\Omega+i\varepsilon}{q_1}+
\Big[\Big(\dfrac{\Omega+i \varepsilon}{q_1}\Big)^2-1\Big]
\ln\dfrac{(\Omega+i\varepsilon)/q_1-1}{(\Omega+i\varepsilon)/q_1+1}\Bigg\}.
$$

Longitudinal dielectric permeability of degenerate plasmas is equal to:
$$
\varepsilon_l=1+\dfrac{3}{q_1^2}
\dfrac{1+\dfrac{\Omega+i\varepsilon}{2q_1}
\ln\dfrac{\Omega+i\varepsilon-q_1}{\Omega+i\varepsilon+
q_1}}{1+\dfrac{i\varepsilon}{2q_1}
\ln\dfrac{\Omega+i\varepsilon-q_1}{\Omega+i\varepsilon+q_1}}.
$$

If to introduce the variable
$$
Z=\dfrac{\Omega+i\varepsilon}{q_1},
$$
that expressions for transversal and longitudinal permeability will write down
more shortly:
$$
\varepsilon_{tr}=1-\dfrac{3}{4\Omega q_1}
\Big[2Z+(Z^2-1)\ln\dfrac{Z-1}{Z+1}\Big],
$$
$$
\varepsilon_l=1+\dfrac{3}{q_1^2}\Big[1+\dfrac{Z}{2}\ln\dfrac{Z-1}{Z+1}\Big]
\Big[1+\dfrac{i\varepsilon}{2q_1}
\ln\dfrac{Z-1}{Z+1}\Big]^{-1}.
$$

Let's transform now functions $Z_1$ and $Z_2$. For the first function
according to (5) it is received:
$$
Z_{1}=-\dfrac{4i}{\omega d^3}\sum\limits_{n=1}^{+\infty}
\dfrac{1}{q^2_{2n-1}}
\Bigg[\dfrac{\pi^2(2n-1)^2}{\varepsilon_{tr}(q_{1,2n-1})-
c^2q^2_{2n-1}/\omega^2}+
\dfrac{(kd)^2}{\varepsilon_l(q_{1,2n+1})}\Bigg],
$$
where
$$
q_{1,2n+1}=\dfrac{v_F}{\omega_p}q_{2n-1}, \qquad
q_{2n-1}=\sqrt{\dfrac{\pi^2}{d^2}(2n+1)^2+k^2}.
$$

For the second function according to (5) it is received following expression
$$
Z_2=-\dfrac{2ick^2}{\omega d\varepsilon_l(q_{1,0})}-\dfrac{4ic}{\omega d^3}
\sum\limits_{n=1}^{\infty}\dfrac{1}{q_{2n}^2}\Bigg[
\dfrac{\pi^2(2n)^2}{\varepsilon_{tr}(q_{1,2n})-c^2q_{2n}^2/\omega^2}+
\dfrac{(kd)^2}{\varepsilon_l(q_{2n})}\Bigg],
$$
where $$
q_{2n}=\sqrt{\dfrac{\pi^2}{d^2}(2n)^2+k^2},\qquad
q_{1,2n}=\dfrac{v_F}{\omega_p}q_{2n}.
$$

\begin{center}
{\bf 3. Surface plasmon in case of symmetry on a magnetic field}
\end{center}

From the third equation of system (3) it is received the following
com\-mu\-ni\-ca\-tion between
projections electric and magnetic fields nearby from the bottom surface
of a layer out of it (when $j_z=0$)
$$
\alpha H_y(0)=-\dfrac{i\omega}{c}E_z(0),
$$
whence for the impedance we obtain the expression
$$
Z_j=\dfrac{i\alpha c}{\omega},\qquad j=1,2.
\eqno{(6)}
$$

Considering the equation (2) expression (6) can be transformed to the form
$$
\sqrt{k^2-\dfrac{\omega^2}{c^2}}=-\dfrac{i\omega
}{c}Z_j,\qquad j=1,2.
\eqno{(7)}
$$

For thin films and small $k$ earlier we had obtained the following
expression
\cite{latyush2011} $\omega^2=4(ck)^2/(4+k^2d^2)$.

Within the limits of macroscopical electrodynamics for great values
of wave number $k$ for frequency of surface plasma oscillations
fairly following equation \cite{R1988}
$$
\omega^\pm(k)=\sqrt{\dfrac{1\pm e^{-kd}}{2}}\,\omega_p.
$$

Let's consider the case when the film consists of a layer of potassium.
Then \cite{F69}$\omega_p=6.5\times 10^{15} sec^{-1}$,
$v_F=8.52\times 10^7$ cm/sec.

On fig. 1--6 we will present dependence of the real and imaginary
parts of frequency plasma oscillations from wave number for films with
the thickness equals to 5, 10 and 50 nanometers.

From drawings 1--3 it is visible, that for thin films
(in particular, with the thickness equals to 5 and 10 nm) the real part
of frequency of plasma oscillations has the maximum near to wave number $k=1$.
Thus at $0<k<1$ there is a sharp increase of the real part of frequency of
oscillations, at $k>1$ goes its smooth decrease.
    With growth of a thickness of a
film the maximum of the real part frequencies of plasma oscillations
is erased. The quantity $\Re\Omega(k)$ becomes
monotonously increasing.

On fig. 4--6 dependence of an imaginary part of
frequency of plasma oscillations from wave number is presented.
    Irrespective of a thickness
films this dependence is monotonously decreasing.
    Near to the wave
numbers $k=1$, to be exact, at $0 <k <1$ occurs sharp decrease of the imaginary
parts of frequency of plasma oscillations, and at $k>1$ such decrease
is smooth.
    With growth of a thickness of a film sharp decrease near to a point
$k=1$ smoothes out.

\begin{figure}[htb]\center
\includegraphics[width=15.0cm, height=10cm]{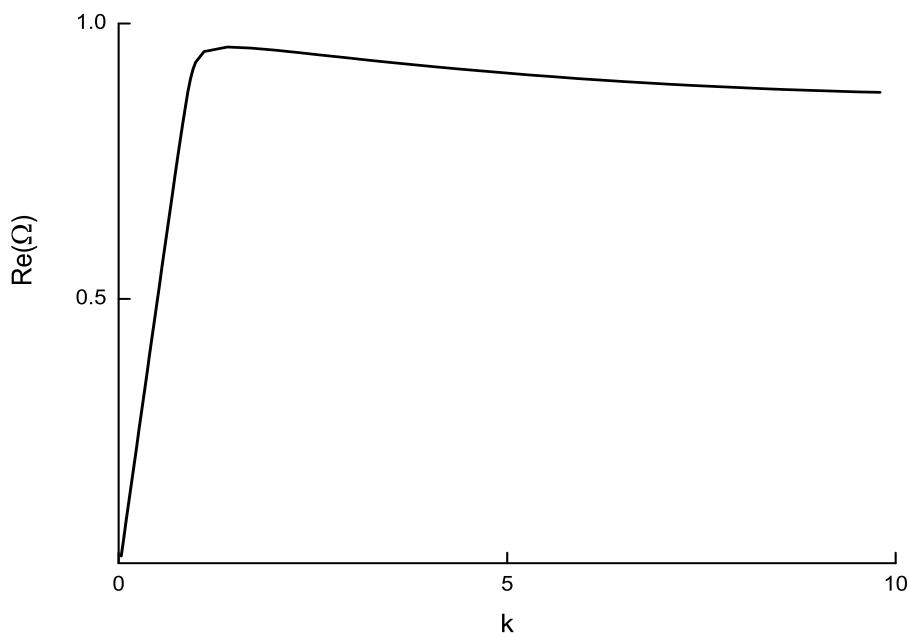}%
\noindent\caption{Surface plasmon. The thickness of film equals $5$ nm.
The dependence $\Re\Omega(k)$.} \end{figure}

\begin{figure}[htb]\center \includegraphics[width=15.0cm, height=10cm]{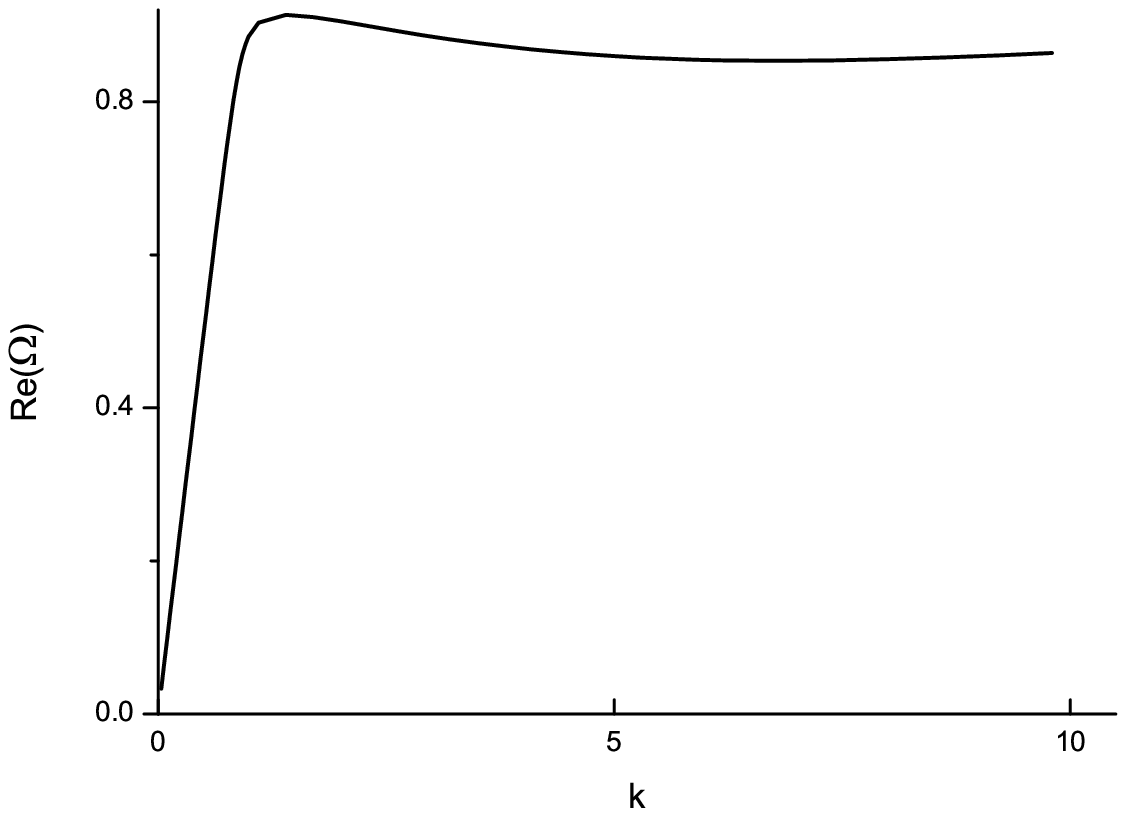}%
\noindent\caption{Surface plasmon. The thickness of film equals $10$ nm.
The dependence $\Re\Omega(k)$.} \end{figure}

\begin{figure}[htb]\center
\includegraphics[width=16.0cm, height=10cm]{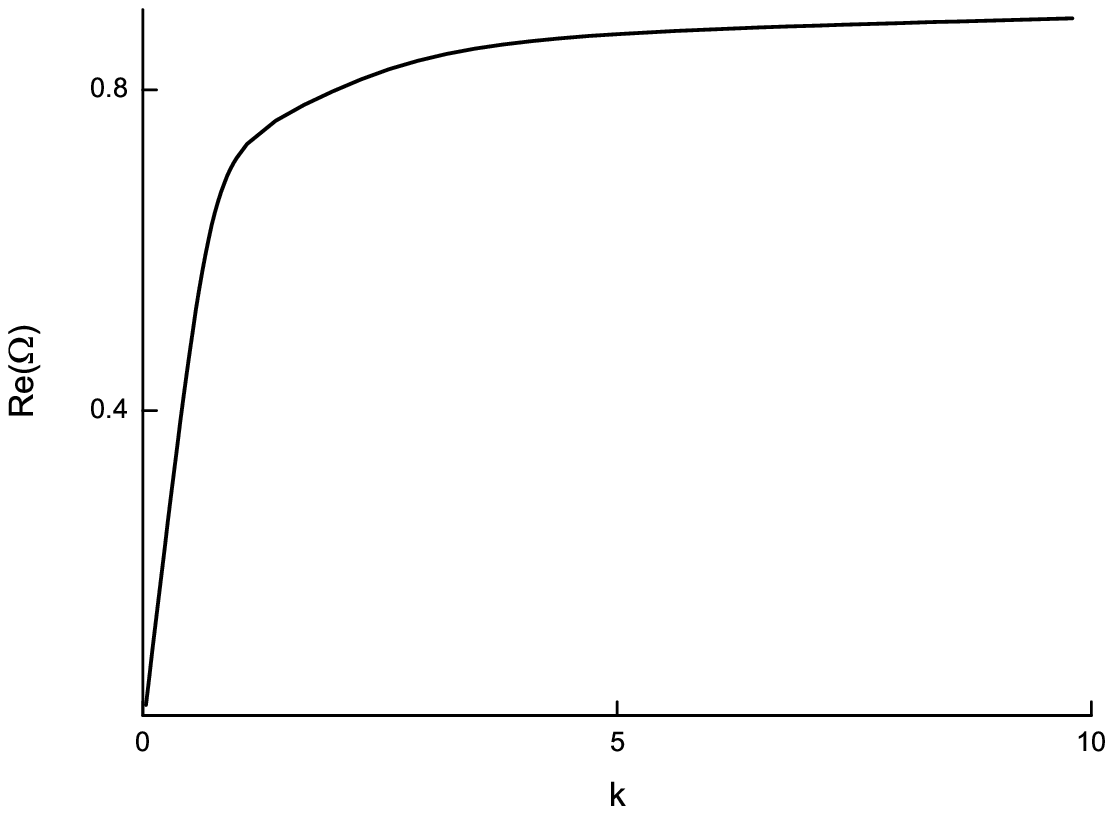}
\noindent\caption{Surface plasmon. The thickness of film equals $50$ nm.
The dependence $\Re\Omega(k)$.} \end{figure}

\begin{figure}[htb]\center
\includegraphics[width=15.0cm, height=10cm]{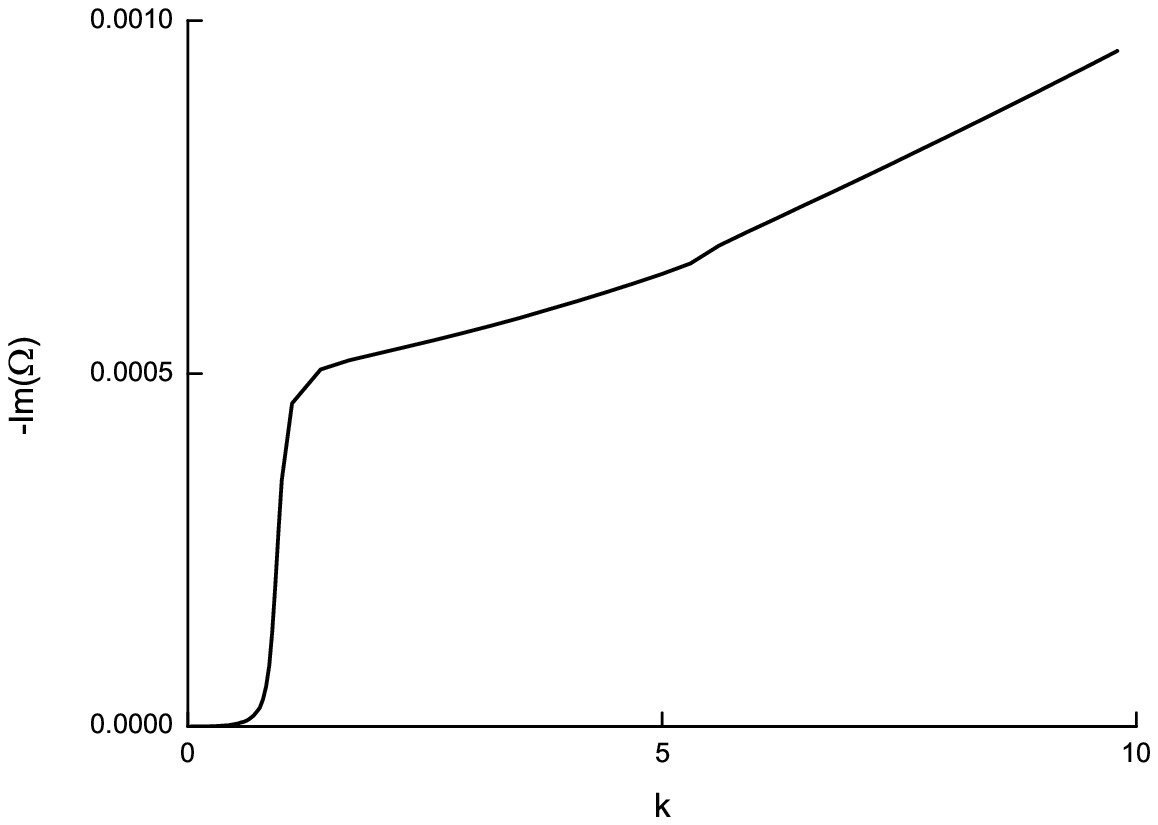}%
\noindent\caption{Surface plasmon. The thickness of film equals $5$ nm.
The dependence $-\Im\Omega(k)$.} \end{figure}

\begin{figure}[htb]\center
\includegraphics[width=15.0cm, height=10cm]{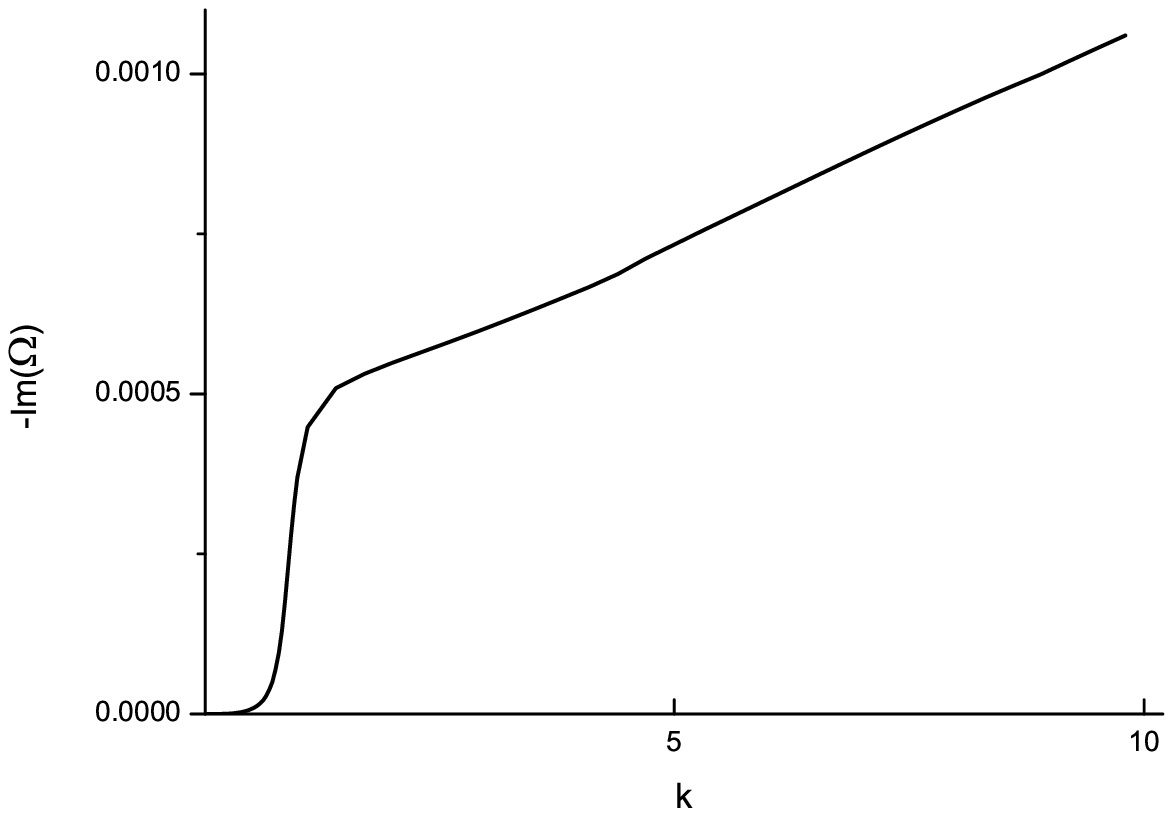}%
\noindent\caption{Surface plasmon. The thickness of film equals $10$ nm.
The dependence $-\Im\Omega(k)$.}
\end{figure}

\begin{figure}[htb]\center
\includegraphics[width=16.0cm, height=10cm]{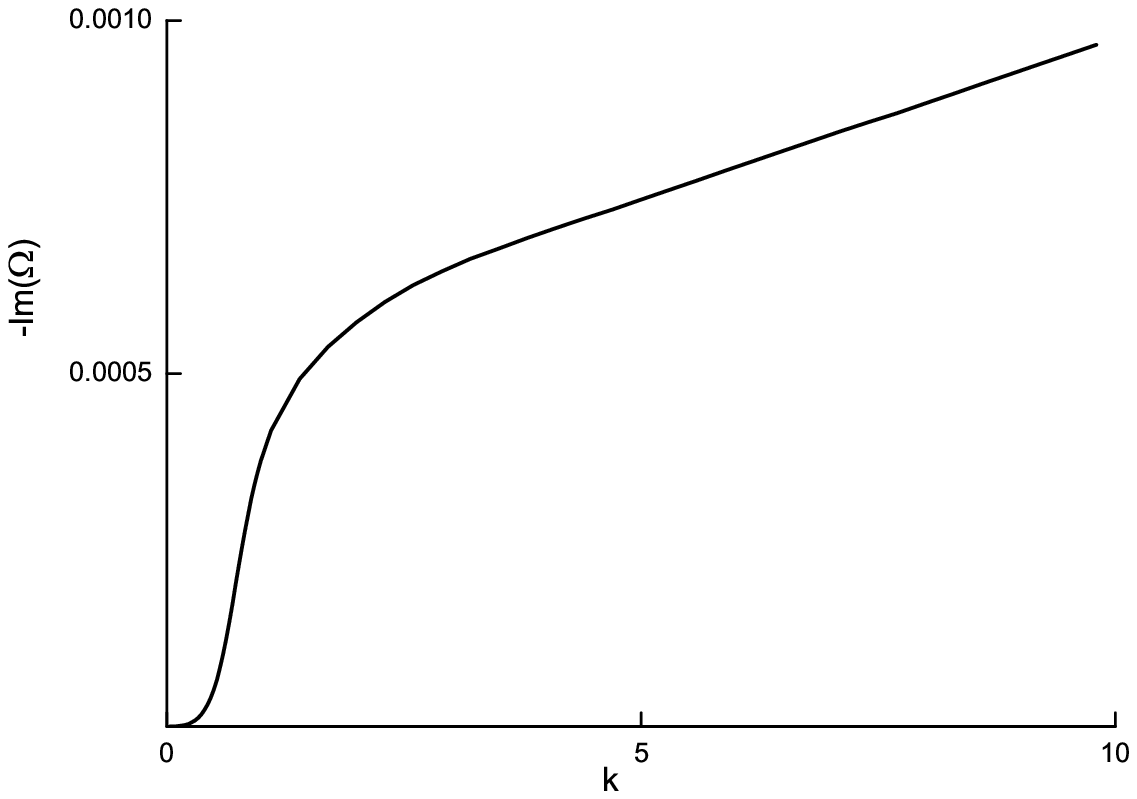}
\noindent\caption{Surface plasmon. The thickness of film equals $50$ nm.
The dependence $-\Im\Omega(k)$.} \end{figure}

\end{document}